# Episodic formation of cometary material in the outburst of a solar-like young star

P. Ábrahám[1], A. Juhász[2], C.P. Dullemond[2], Á. Kóspál[3], R. van Boekel[2], J. Bouwman[2], Th. Henning[2], A. Moór[1], L. Mosoni[1,2], A. Sicilia-Aguilar[2], N. Sipos[1]

[1]*Konkoly Observatory of the Hungarian Academy of Sciences, P.O. Box 67, 1525 Budapest, Hungary.* [2]*Max-Planck-Institut für Astronomie, Königstuhl 17, 69117 Heidelberg, Germany.* [3]*Leiden Observatory, Niels Bohrweg 2, 2333 CA Leiden, The Netherlands.*

**Our Solar System originated in interstellar gas and dust; the latter is in the form of amorphous silicate particles[1,2] and carbonaceous dust. The composition of cometary material shows that a significant fraction of the amorphous silicates was transformed into crystalline form during the early evolution of the protosolar nebula[3]. How and when this transformation happened has been controversial, with the main options being heating by the young Sun[4,5] or shock heating[6]. Here we report mid-infrared features in the outburst spectrum of the young solar-like star EX Lupi that were not present in quiescence. We attribute them to crystalline forsterite; the crystals were produced via thermal annealing in the surface layer of the inner disk by heat from the outburst, a process that has hitherto not been considered. The observed lack of cold crystals excludes shock heating at larger radii.**

The year 2008 brought a rare opportunity to study high temperature dust processing on human timescale in a cosmic laboratory. The experiment took place in the circumstellar disk of EX Lupi, an M0 star, that is the prototype of a class of young eruptive stars named EXors[7]. These objects are defined by their large repetitive outbursts, attributed to temporarily increased mass accretion from the circumstellar disk onto the star[8]. Such outbursts represent the most intense accretion episodes in assembling the final stellar mass. In January 2008 EX Lupi entered one of its largest outbursts, brightening by approximately a factor of 100 and reaching a maximum brightness of 8 magnitude in visual light[9,10]. We observed EX Lupi in the 5.2–37 μm wavelength range with the InfraRed Spectrograph on-board the Spitzer Space Telescope, on 2008 April 21. EX Lupi was already slowly fading after its peak brightness in 2008 February, but was still a factor of 30 brighter in visual light than in quiescence.

Comparing our EX Lupi spectrum with a pre-outburst measurement from the Spitzer archive obtained in 2005, we observed a remarkable change. In the 8 – 12 μm spectral range (Figure 1), the silicate profile in quiescence exhibited a triangular shape, similar to that of the amorphous interstellar grains[1,11] (panels a, b). In contrast, in outburst (panel c), we clearly detected several narrower spectral features, which we identified as crystalline silicates, on top of the broad peak of amorphous silicates. The sharp peak at 10 μm and a shoulder at 11.3 μm suggest that forsterite, the Mg-rich form of olivine, dominates the observed crystal population[12,13]. The appearance of a weaker peak at 16 μm, shown in the Supplementary Information, supports this conclusion. Longward of this peak no other crystalline features are present in the spectrum. The observed crystalline features are similar to those present in comet spectra[3,14] (panel d) and in a number of protoplanetary disks[15,16]. The remaining differences between EX Lupi and the cometary spectra can be related to different temperatures and different relative abundances of dust components.

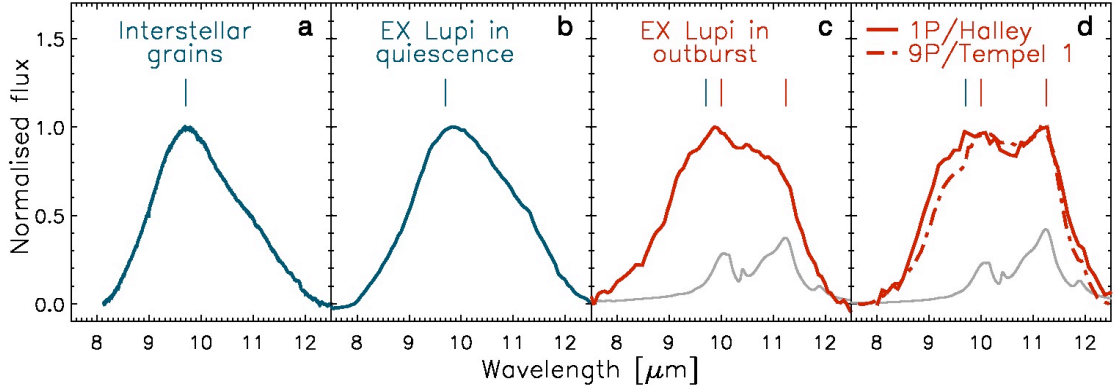

**Figure 1** Silicate emission in the 8 – 12 μm range. (a) Spectrum of interstellar grains measured in the direction of the Galactic Center[1]; (b) Spitzer IRS spectrum of EX Lup obtained on 2005 March 18 in quiescent phase; (c) our Spitzer spectrum of EX Lup obtained on 2008 April 21, in the middle of the present outburst; (d) solid line: ground-based spectrum of Comet 1P/Halley[3], dash-dot line: Spitzer spectrum of the ejecta from Comet 9P/Tempel 1 during the Deep Impact experiment[14] (available in the Spitzer archive). After a linear continuum removal, the spectra were normalized to their peak value. Panel (a) exhibits the characteristic triangular shape profile attributed to amorphous silicate grains[1]; the vertical blue dash at 9.7 μm (repeated in all panels) corresponds to the peak wavelength of the amorphous silicate profile as measured in laboratory[11]. In panel (b) the EX Lupi spectrum closely resembles the amorphous profile, with some slight excess on the long wavelength side. In panel (c) peaks and shoulders due to crystalline silicates can be identified. Peak wavelengths of forsterite at 10.0 and 11.2 μm, as measured in laboratory experiments[12,13], are marked by red dashes. The solid gray curves in panels (c) and (d) display the emissivity curve of pure forsterite[13] assuming representative silicate grain temperatures of 1250 K and 300 K, respectively. Panel (d) shows that the same crystalline features can be observed in cometary spectra.

Since the quiescent spectrum exhibits no crystalline silicate features, the appearance of crystalline features in the EX Lupi outburst strongly suggests that we witness on-going crystal formation; to our knowledge for the first time in a celestial object. Alternative explanations, such as illumination of existing crystals residing in outer disk areas or stirring up crystals from the disk mid-plane can be excluded via modeling, detailed in the Supplementary Information accompanying this paper. The relative strengths of the 10 and 11.3 μm crystalline features, and the lack of spectral features beyond 16 μm imply that the new crystals are hot, and were formed in a high temperature process. We estimated a temperature range for this process utilizing our recent radiative transfer modeling of EX Lupi in quiescence[17], which assumes a circumstellar disk geometry encircling an inner dust-free hole of 0.2 AU (see Supplementary Information, Sect. 1.3). Due to this hole, the temperature in the quiescent disk was almost everywhere below 700 K. The quiescent spectrum indicates that no noticeable crystal formation has occurred at these temperatures. The disk temperature in outburst was simulated by increasing the luminosity of the central source by a factor of 10, estimated from the flux increase at the highest frequency part of the Spitzer spectra at around 5 μm. An assumed black-body spectrum of 6800 K, typical of outbursting stars[8], accounted for the higher temperature of the source during eruption. The modeling revealed that in outburst a significant disk area became hotter than 700 K, but its temperature was almost everywhere below 1500 K, the vaporization threshold of silicate particles. Our observations point to a crystallization mechanism which works efficiently between 700 and 1500 K in the protoplanetary environment. Laboratory experiments suggest that this mechanism is thermal annealing[18,19]. According to laboratory measurements, above 1000 K annealing occurs on very short timescales of seconds to hours[18], fitting well within the observed timescale of the EX Lupi outburst. The radius of this crystal formation zone, 0.5 AU, is comparable to the terrestrial planet region in our Solar System. The high temperature of the crystals excludes their formation at several AU radii in shock fronts[6], because grains behind the shock would quickly cool down and produce observable spectral features longward of 20 μm.

Our observations revealed another interesting process acting in the surface layer of the disk. In 1955-56 EX Lupi had already a major eruption very similar to the present one[7], in which probably a similar amount of crystalline silicates was formed (see Supplementary Information, Sect. 1.4). However, by 2005, the date of the quiescent Spitzer spectrum, the crystalline spectral features vanished, indicating that the crystals disappeared from the disk surface layer in less than 50 years. Possible explanations for this fast removal process, discussed more in the Supplementary Information, are vertical mixing which transports the crystals into the disk interior[4,20], inward surface flow which may accrete them onto the star[20]; or amorphization by cosmic rays[21] or X-rays[22]. Note that in the similar, but even more violent eruptions of FU Orionis-type young stars crystals were searched for but not detected, and their lack was speculated to be the effect of vertical mixing[23].

Our detections demonstrate that crystalline silicate grains can both form in and disappear from the surface layer of the disk within months to decades. Thus we predict that multi-epoch measurements of the crystallinity level in EX Lupi would provide fluctuating results, depending on the actual activity level of the star and in particular on the time elapsed since its last major outburst. Therefore, the observed crystallinity is not a useful indicator of the dust composition in the disk interior, which would evolve much more gradually. Also, the observed dominance of forsterite may not be representative of the disk interior. Similar conclusions may hold for many other young stars. Pre-main sequence evolution is generally accompanied by optical-infrared variability[24,25], and a large fraction – if not the majority – of young stars frequently change their luminosity by factors of < 10. Thus the observed crystallinity may vary considerably – and randomly – among stars of similar mass and age, and correlation of its value with the stellar parameters may be weaker and less informative than previously expected.

The observations of the outburst of EX Lupi point to a new scenario for crystal formation in protoplanetary disks: episodic surface crystallization. It is now an observationally established crystal forming mechanism acting in the inner disk region, supplementarily to crystal formation due to accretion heat in the disk mid-plane[4] or shocks[6]. We suggest that such crystallization events occur in the life of most young stars. With an average interval of 50 years, EX Lupi may undergo several thousand such eruptions during its early evolution. The current paradigm confines crystallization to the very early phases of pre-main sequence evolution[4]. Our findings show that crystallization episodes can also continue in later phases when accretion already significantly dropped (apart from the episodic outburst), covering a significant part of the pre-main sequence evolution. Moreover, this scenario may work even in disks with large inner holes (like EX Lup), where the mid-plane temperature will never be high enough for crystallization. Our proposal on episodic crystallization is in line with recent observations which imply that there must be another grain crystallizing mechanism uncorrelated with the steady mass accretion rate, stellar luminosity, disk mass, or disk/star mass ratio[26]. Although in a single outburst only the thin surface layer of the disk is crystallized, in EX Lupi we saw that subsequent major outbursts transform always a new layer of amorphous grains to crystals, potentially enriching the disk interior via vertical mixing. In any case, a fraction of the crystals may be mixed outwards, and contribute to the build-up of proto-comets. Assuming that a similar process worked in the proto-Solar System, crystalline grains in comets and meteorites might be messengers of past eruptions, having been formed in a melting pot around the outbursting young Sun.

**Acknowledgements** We are grateful to Albert F. Jones for providing us with timely visual observations of EX Lupi during the preparation of our infrared measurements. The presented work was partly supported by the Hungarian Research Fund (OTKA). The research of Á. K is supported by the Netherlands Organization for Scientific Research (NWO).

# Supplementary Information

# Supplementary Discussion

Spitzer observations and data reduction

We obtained spectra of EX Lupi with the Infrared Spectrograph (IRS)[27] on-board the Spitzer Space Telescope, on 2008 April 21. We used both the low-resolution modules (R = 60−120) in the 5.2–14.5 μm wavelength range, and the high-resolution modules (R = 600) between 9.9–37.2 μm. A high accuracy PCRS peak-up was executed prior to the spectroscopic observations to position the target within the slit. The integration time was set to six seconds for all modules, with four observing cycles for redundancy. For the low-resolution observations, the background was subtracted using associated pairs of imaged spectra from the two nod positions along the slit, also eliminating stray light contamination and anomalous dark currents. For the high resolution observations a separate background measurement was performed with the same set-up.

Pixels flagged by the data pipeline as being "bad" were replaced with a value interpolated from an 8 pixel perimeter surrounding the errant pixel. Our spectra are based on the droopres and rsc products processed through the S15.3.0 version of the Spitzer data pipeline for the low and high resolution data, respectively. The low resolution spectra were extracted using a 6.0 pixel fixed width aperture in the spatial dimension, while for the high resolution channels we extracted the spectra by fitting the source profile with the known psf in the spectral images. The low-level fringing in the high resolution spectra was removed using the irsfringe package[28]. The spectra are calibrated using a spectral response function derived from IRS spectra and MARCS stellar models for a suite of calibrators provided by the Spitzer Science Centre. To remove any effect of pointing offsets, we matched orders based on the point spread function of the IRS instrument, correcting for possible flux losses. A section of the spectrum centred on the forsterite peak at 16 μm is shown in Fig. 1.

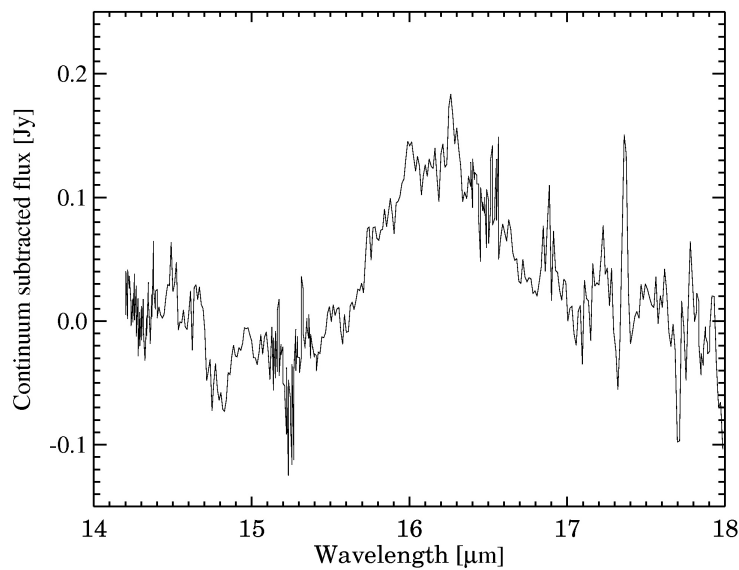

**Figure 1:** The crystalline silicate peak at 16 μm in the outburst spectrum. Continuum fluxes in the ranges of 14.2–15.0 μm and 17.2–18.0 μm were fitted by a second order polynomial and subtracted from the data.

A pre-outburst observation was obtained by Stringfellow et al. (PID: 3716) on 2005 March 18 which is now available in the Spitzer archive. The instrumental set-up was identical with our outburst measurement, except that the total exposure time was longer, and no separate background measurement for the high resolution module was obtained. The pre-outburst spectra were extracted in an identical way as our outburst spectra, with the exception that for the high-resolution spectra the background emission was removed by fitting a local continuum underneath the source profile.

## Alternative scenarios to on-going crystal formation

One may wonder which other possible scenarios, apart from formation of new silicate crystals, could have produced the observed peaks in the outburst Spitzer IRS spectrum. We investigated two alternative scenarios, both of which assume that crystals already existed in the disk before the outburst and they were not created in the recent eruption.

In the first scenario the crystals are located in the disk midplane, where they were created earlier due to viscous heating. These crystals were stirred up to the disk atmosphere in the 2008 outburst, where they became visible for infrared spectroscopy. We studied the validity of this scenario by calculating the turbulent mixing timescale required to move a dust grain from the disk midplane to the disk atmosphere. It can be seen in Fig. 2 that the resulting timescales are longer, even for the strongest possible turbulence, than the approximately three months elapsed between the beginning of the outburst and the Spitzer IRS measurement.

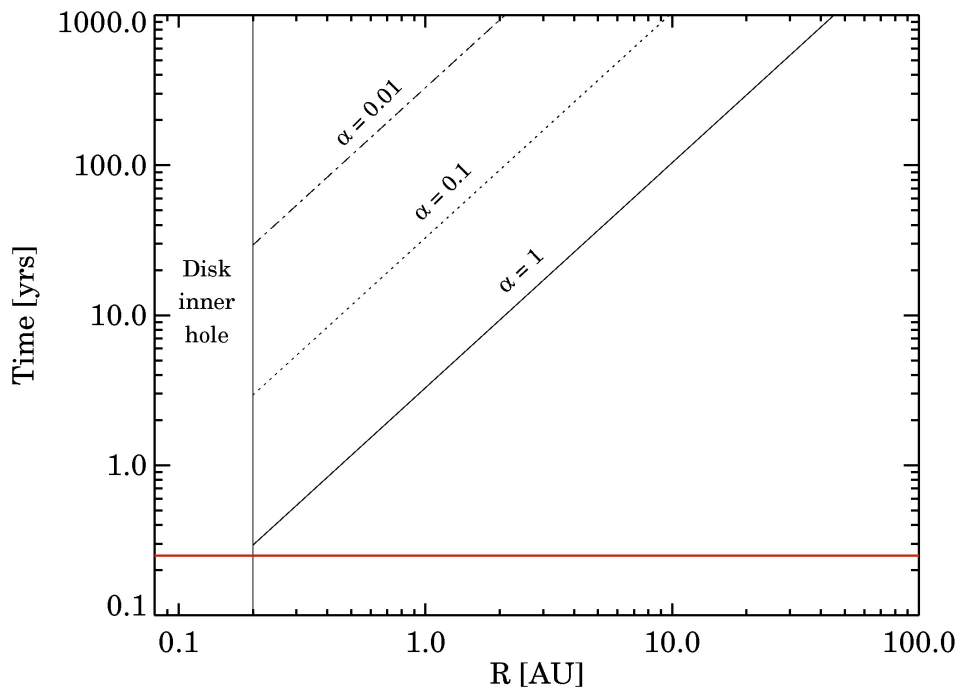

**Figure 2:** Timescale for mixing up a dust particle vertically from the mid-plane to the disk atmosphere, as a function of the distance to the central star in the plane of the disk, for different values of the turbulent viscosity parameter α. The horizontal red line is set to 3 months, the time elapsed between the beginning of the outburst and our Spitzer IRS measurement.

The other scenario assumes that the existing crystals were already located in the disk atmosphere in the outer disk and the outburst only illuminated and not created them. In the quiescence the temperature of the crystals should be low enough to prevent the crystals to produce mid-infrared features in the IRS wavelength range (5.5–38 μm). In order to avoid the appearance of the strong 33 μm band, the temperature of the silicate crystals should be lower than 100 K. In the outburst the temperature will rise in the whole disk due to the enhanced irradiation luminosity, making the crystals hot enough to produce mid-infrared features. Since in the outburst spectrum we do not see any crystalline feature longwards of 16 μm, the crystals should be very hot (>700 K) in order to avoid the appearance of the strong bands at 24 μm and at 33 μm. If the above described illumination effect would be the explanation of the difference between the outburst and quiescent phase spectra, the temperature of the crystals should change by more than a factor of 6. Since the temperature in the disk is proportional to the square-root of the irradiation luminosity, this temperature difference indicates a luminosity change of more than a factor of 36. This is by far above the observed change in the infrared luminosity, which is directly proportional to the irradiation luminosity. Thus in our view the only possible scenario is that crystals are formed in-situ during the outburst in the upper layer of the disk of EX Lup.

The EX Lupi system

EX Lupi is an M0V star at a distance of about 155 pc. According to our modeling[17], performed with the radiative transfer code RADMC[29], the star harbours a circumstellar disk which is passively heated by the radiation of the star. Accretion heating in the mid-plane of the disk turned out to be insignificant both in quiescence and in outburst. The geometry of the disk is depicted in Fig. 3. It is a moderately flaring disk with a mass of 0.025 $M_\odot$. The inner dust-free hole has a radius of 0.2 AU, which is larger than the sublimation radius of silicate grains (0.05 AU), thus the origin of this large hole is unclear.

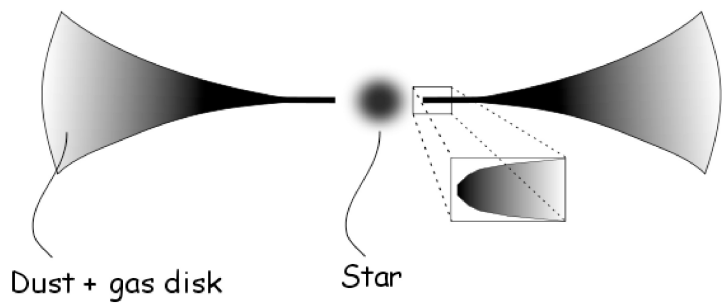

**Figure 3:** Schematic picture of the geometrical structure of the EX Lupi system in quiescent phase.

The radial temperature profiles of the disk in quiescence and in outburst are shown in Fig. 4. In quiescence the surface temperature was almost everywhere below 700 K, only in a narrow ring located at the inner edge of the disk (0.2 < r < 0.21 AU) increased the temperature to T ≤ 900 K. In outburst the whole disk became hotter and the inner 0.5 AU area was heated above 1000 K, making possible crystal formation. The temperature nowhere exceeded 1500 K, the sublimation threshold, except at the inner edge of the disk where it reached a peak temperature of 1700 K. The mid-plane remained below the crystallization temperature both in quiescence and outburst.

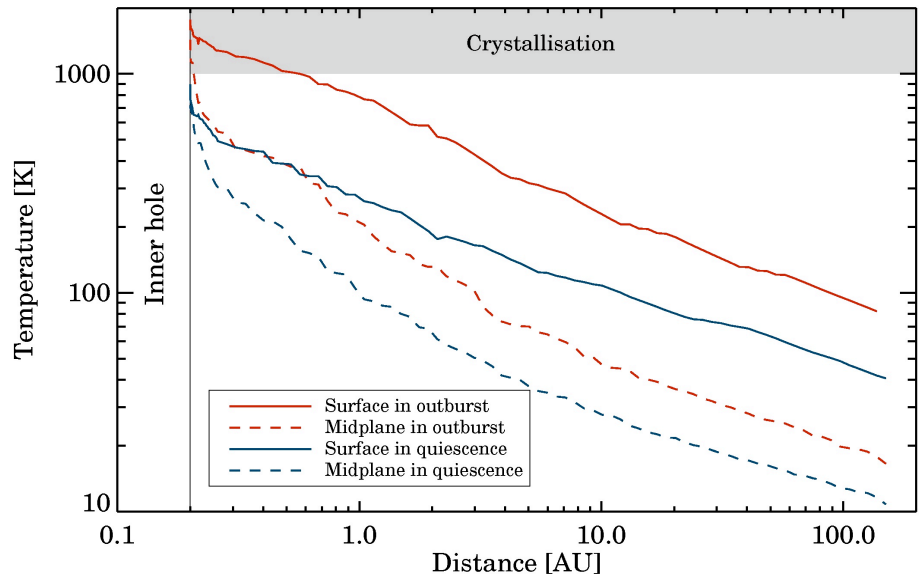

Figure 4: Radial temperature profile of the disk surface (continuous line) and the disk mid-plane (dashed line) in quiescent phase (blue) and in outburst (red). The radius of the inner dust-free hole and the temperature range where crystal formation may take place (T > 1000K) are marked.

## The outburst history

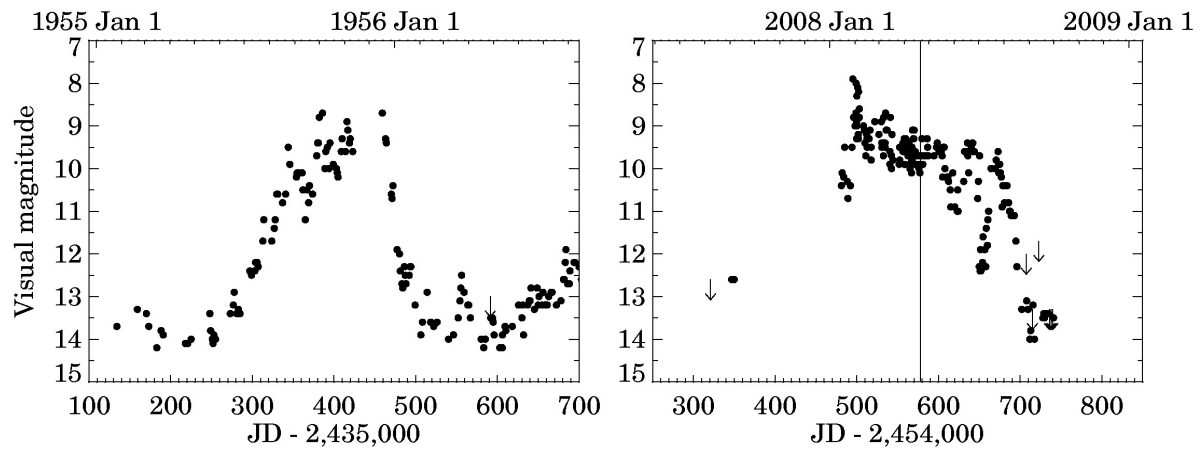

**Figure 5:** Light curve of EX Lup compiled from visual estimates. Left: the outburst of EX Lupi in 1955-56. Right: the recent outburst in 2008, taken from data at www.aavso.org. The peak brightnesses and the characteristic timescales were similar in the two eruptions. Our Spitzer observation was obtained on 2008 April 21, marked by the vertical line in the figure.

The present outburst was not the first one in the life of EX Lupi. Beside smaller scale flarings by factors <10, it had a major eruption in 1955-56, when the peak visual brightness reached 8.5 magnitude, very close to the maximum of the present outburst[7] (Fig. 5). The comparable peak luminosities in the two eruptions imply that the disk areas heated above the crystallization threshold were also similar. Since crystallization above 1000 K is an almost instantaneous process, it is the maximum illuminated area, rather than the duration of the brightness peak, which determines the crystal yield. It is then reasonable to assume that very similar amounts of crystalline silicate were formed in the two eruptions, and the shape of the silicate profile in early 1956 would have looked similar to the one we measured in 2008 April.

The disappearance of crystals after outburst

Because the silicate profile in early 1956 should have probably looked similar to the one measured in 2008, it is surprising that by 2005, the date of the quiescent Spitzer spectrum, the crystalline spectral features vanished. Possible explanations are vertical mixing in the disk which transports the crystals into the disk interior; inward surface flow which may accrete them onto the star; or amorphization of crystals by cosmic rays or X-rays. Exploring the first scenario, we computed time-scales for vertical mixing of grains over the width of the surface layer (the disk interior is probably mainly amorphous, as indicated by the quiescent spectrum), for different values of $\alpha$, the viscosity parameter in the standard alpha-prescription of disks (Fig. 6). On theoretical basis, the value of $\alpha$ lies between $10^{-5}$ and 1, higher values corresponding to higher levels of turbulence. Though its value is fundamental in hydrodynamic simulations of planet formation, only very weak observational constraints exist. Adopting the 50 year interval between the two major outbursts of EX Lupi as an upper limit for crystal removal, Fig. 6 constrains $\alpha$ to be higher than $10^{-4}$. This is a plausible result, indicating that vertical mixing can be an explanation for the quick removal of crystals. The inward surface flow scenario might also explain the disappearance of crystals. Surface flows are predicted in two-dimensional transport models of viscous protoplanetary disks[20,30]. Based on these simulations we estimated that the inward surface velocity is approximately 0.0006 AU/yr for the gas component. Assuming that the crystals are coupled to the gas, in 50 years they could proceed 0.03 AU, one order of magnitude less than the size of the crystallized zone of 0.5 AU in EX Lupi. This estimate, however, depends on the exact choice of $\alpha$ (assumed to be 0.001 in the mentioned simulations), and higher $\alpha$ values would result in higher surface flow velocities potentially able to accrete the fresh crystals onto the star within 50 years. Note, however, that 50 years is only an upper limit for crystal removal, and also that in detailed modeling the differences between gas and dust velocities have to be taken into account.

On the variability of young stars

Photometric variability is typical of most pre-main sequence stars with and without disks[24,31]. Accretion is a very dynamic process in classical T Tauri stars (CTTS)[32], being one of the leading causes of strong, non-periodic variations. Accretion-related hot spots, together with variable extinction caused by circumstellar material in nearly edge-on systems, cause photometric variations in most of the young CTTS[24]. Both variability mechanisms can be distinguished by observing the colour changes and accretion indicators (e.g. accretion-related emission lines). Photometric monitoring of large samples of CTTS over years/decades[24,33] reveals that more than 50% of the CTTS with ages 1-3 Myr exhibit variations over 1 magnitude in the V band, and 10% of them experience even larger variations ($\Delta V \leq$ 1.5-2 mag). Although objects with low-amplitude variations display typical variability patterns that change little during years/decades, the most extreme variables are typically erratic[33], so stars with large variations also have epochs with lower-amplitude variability. In addition, the amplitude of the variations measured is always a lower limit, since the light curve changes can be very fast. Therefore, accretion variations and small accretion outbursts may be common for a large fraction of pre-main sequence stars during their first Myrs of life.

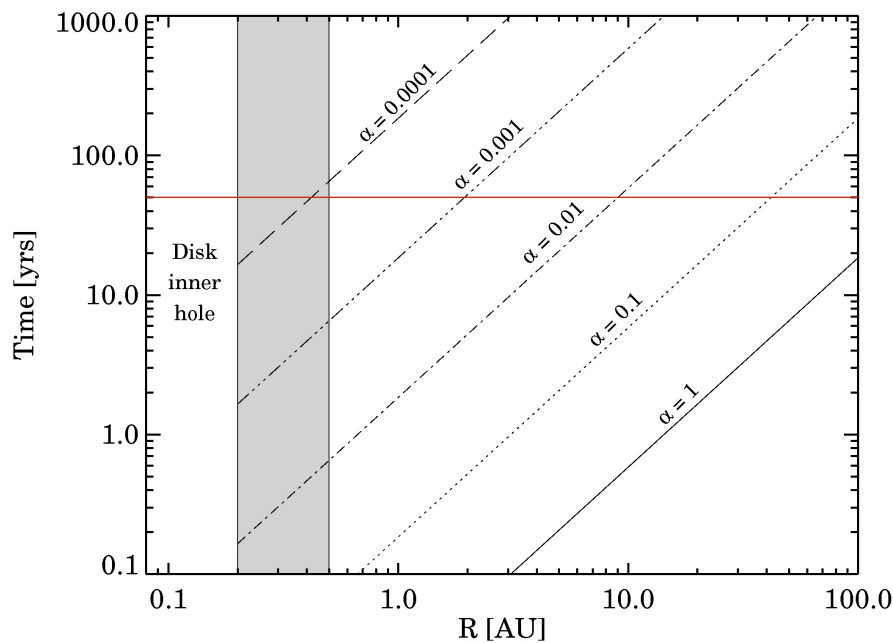

**Figure 6:** Vertical mixing timescale as a function of the distance to the central star in the plane of the disk, for different values of the turbulent viscosity parameter α. The shaded area marks the disk region where the temperature is sufficiently high for crystallization during outburst. The horizontal red line is set to 50 years, the time elapsed since the last outburst of similar magnitude.

## Supplementary Notes

27. Houck, J. R., Roellig, T. L., van Cleve, J. et al. The Infrared Spectrograph (IRS) on the Spitzer Space Telescope. Astrophys. J. Suppl. Ser. 154, 18-24 (2004)

28. Lahuis, F. & Boogert, A. How to Get Rid of Fringes in SIRTF/IRS Data. SFChem 2002: Chemistry as a Diagnostic of Star Formation (Proceedings of a conference held August 21-23, 2002 at University of Waterloo, Waterloo, Ontario, Canada, ed. C. L. Curry & M. Fich), 335 (2003)

29. Dullemond, C. P., & Dominik, C. Flaring vs. self-shadowed disks: The SEDs of Herbig Ae/Be stars. Astron. & Astrophys. 417, 159-168 (2004).